\documentclass[aps, twocolumn, showpacs, letterpaper]{revtex4}

\usepackage{color}
\usepackage{amsmath}
\usepackage{amssymb}
\usepackage{graphicx}
\usepackage{xspace}

\newcommand{\Eq}[1]{Eq.~(\ref{#1})}

\begin{document}
\title{Non-equilibrium piezo-thermal effect in spinning gas}

\author{V.~I. Geyko and N.~J. Fisch}
\affiliation{Department of Astrophysical Sciences, Princeton University, Princeton, New Jersey 08544, USA}

\begin{abstract}
A spinning gas, heated adiabatically through axial compression, is known to exhibit a rotation-dependent heat capacity. 
However, as equilibrium is approached, a new effect is identified here wherein the temperature does not grow homogeneously in the radial direction, but develops a temperature differential with the hottest region on axis,  at the maximum of the centrifugal potential energy. 
This phenomenon, which we call a {\it piezo-thermal effect}, is shown to  grow bilinearly with the compression rate and the amplitude of the potential. 
Numerical simulations confirm a simple model of this effect, which can be generalized to other forms of potential energy and  methods of heating. 

\end{abstract}
\date{\today}

\maketitle

{\it Introduction:} \
A spinning gas  exhibits a rotation-dependent heat capacity \cite{ref:geyko_e}. 
Under axial compression, a cylinder of gas rotating about its axis, will become hotter, but not as hot absent the rotation.
This effect occurs because, as the gas  becomes hotter,  gas  molecules originally flung by centrifugal forces to the periphery become more homogeneously distributed.
Since the heating changes the moment of inertia, the gas must spin faster to conserve angular momentum. 
Thus, some of the energy expended in compressing the gas goes into to making the gas rotate faster, rather than into heating it, making it softer to compress axially.
The question now posed is: to the extent that the axial compression is not quite infinitely slow, do radial temperature gradients develop?

We identify here that, in fact, radial temperature gradients do develop in the  direction perpendicular to the gas compression.  
By analogy to the piezoelectric effect \cite{ref:arnau}, we term this temperature gradient formation the {\it non-equilibrium piezo-thermal effect}. 
Since the effect  appears only in approaching equilibrium, we emphasize its non-equilibrium feature.
The effect  occurs  in the presence  of any external or self generated potential. 
Simple examples are a gas in a constant gravitational field or a spinning gas in a cylinder, with perfect slip boundary conditions.
With compression  in the direction perpendicular  to the   potential gradient, temperature gradients arise in the direction of the  potential gradient.

Consider first a slab of gas in equilibrium in a vertical gravitational field, so that it has an atmospheric density distribution but constant temperature.  
The force of gravity takes the place of the centrifugal force, but, since the gravitational force is constant, rather than coupled to the temperature,
the essential effect is captured more simply. 
If the slab is compressed slowly in the horizontal direction, it becomes hotter, so the scale height rises.
For mean free path small compared to  system size, at each height  local thermodynamic equilibrium is achieved.
As the gas becomes hotter,  an axial temperature gradient develops,  
with the gas on bottom coldest.
This is the piezo-thermal effect for gas in a gravitational field.

There are several important time scales:  
the  collision time  $\tau_c$, the compression or energy input time $\tau_E$, 
the rise time of the gas or sound time  $\tau_s$, and the heat diffusion time $\tau_H$. 
Consider  the case  $ \tau_c \ll \tau_E \ll  \tau_s  \ll \tau_H$.  
The smallness of $\tau_c$ guarantees local thermodynamic equilibrium.
The largeness of $\tau_H$ allows for temperature differentials.
The inequality $\tau_E \ll  \tau_s$ means ``fast compression," where the energy from compression will first be distributed among all other degrees of freedom, while the density remains  unchanged. 
Although more energy is deposited at the bottom than at the top, the energy deposited per particle is the same throughout the vertical direction, since the energy delivered to the particles is just proportional to the particle number at that height.
Local thermodynamic equilibrium is achieved at increased temperature, but the system  is  not in global equilibrium, so the gas center of mass (COM)   rises to the proper scale height. 
Particles on the bottom then see a receding  COM {\it ceiling}, and hence are  cooled by collisions with the receding ceiling; conversely, particles on top see a rising COM  {\it floor}, and hence are heated.
In the following, these predictions will be described quantitatively and demonstrated through simulations.

Similarly, the axial compression of spinning gas produces a non-equilibrium piezo-thermal effect, with the temperature in the center higher than at the periphery, since the centrifugal force pushes from the center to the periphery. 
In the thermodynamic limit, with adiabatically slow processes, the temperature  and angular velocity  are  constant in space, enabling the derivation of the rotation-dependent heat capacity effect \cite{ref:geyko_e}. 
This effect was exploited to increase theoretical maximum of efficiency of internal combustion engines \cite{ref:geyko_i}. 
However, for engines, the non-equilibrium,  piezo-thermal effects become important, motivating, in fact, the present study.    

\begin{figure}
	\centering
		\includegraphics[scale=0.37, angle=0]{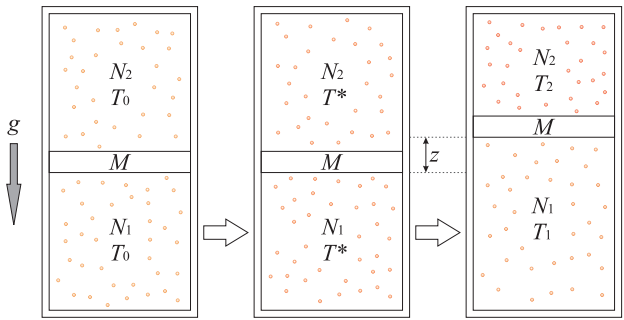}
	\caption{Piezo-thermal effect. Step 1: equilibrium at $T_0$.  Step 2: instant vertically uniform temperature rise to $T^*$ produced by lateral compression.  Step 3:  pressure balance, under heat insulation, restored through partition  displacement $z$.}
	\label{fig:gas_mass}
\end{figure}

{\it Gravitational Potential:} \
To describe quantitatively the piezo-thermal effect,  consider  a slab  of finite width (in the $\hat x$-direction) of an ideal gas in constant vertical gravitational field $-g\hat z$.
In equilibrium, the temperature $T$ is constant, so the density falls off exponentially, $n(z)=n_0 \exp (-z/\lambda)$, where $z$ is the vertical height and   $\lambda = T/mg$ is the atmospheric scale height, for particle mass  $m$.  
Under adiabatic compression  in the $\hat x$-direction, the gas COM  rises, as the temperature increases. 
The increase in the gravitational potential energy causes the gravity-dependent heat capacity of the gas \cite{ref:fai}. 

As will be demonstrated in  particle simulations, the key piezo-thermal effects can be described  in various limits by  crudely modeling the gas, initially at temperature $T_0$, as divided by a movable, heat-insulating  partition with mass $M$ (later  identified with the total gas mass),  with $N_1$  massless molecules in the lower section and $N_2$  in the upper section, as  depicted in Fig.~\ref{fig:gas_mass}.  
Consider first the case  $ \tau_E \ll  \tau_s  \ll \tau_H$, namely where sudden lateral compression  heats the gas in a time $ \tau_E$ short compared to the rise time $ \tau_s $.
The initial  pressure balance  gives
\begin{gather}\label{eq:F_bal}
{N_1 T_0}/{L_1}-{N_2 T_0}/{L_2}=Mg,
\end{gather}
where $L_{1,2}$ are the lengths of the lower and upper sections respectively.  
Upon heating, the temperature increases everywhere from $T_0$ to $T^*$.  
To balance pressures, the partition rises to height $z$, but  gas in the lower section  performs  mechanical work, and hence is cooled, while  gas in the upper section is compressed and heated.  
Since  $ \tau_s  \ll \tau_H$,  the temperatures on top and  bottom  separately equilibrate to $T_2$ and $T_1$ respectively.
For heat  capacity $c_v$  and heat capacity ratio $\gamma$,   energy balance gives
\begin{gather}\label{eq:energy}
c_v(N_1 T^*+N_2 T^*)=c_v(N_1 T_1+N_2 T_2)+Mgz,
\end{gather}
while the adiabatic law for each section gives
\begin{gather}\label {eq:T_X}
T_{1}=T^*\left(1 + {z}/{L_{1}}\right)^{1-\gamma}, \quad T_{2}=T^*\left(1 - {z}/{L_{2}}\right)^{1-\gamma}.
\end{gather}

Now for large scale heights $L\ll \lambda$, we can  approximate $L_{1,2}\approx L/2$, $N_{1,2}\approx N/2$.   For  rough scaling, take the mass of the partition equal to the mass of the gas $M\approx N m$, and expand for small displacements $z/L$, to find
\begin{gather}
\frac{\Delta T}{T_0} =\kappa_1\left(\frac{mgL}{T_0}\right) \left(\frac{T^*-T_0}{T_0}\right),
\label {eq:differential}
\end{gather}
which exhibits  bilinear dependence  on $G=mgL/T_0\equiv L/\lambda$, which measures the system size to scale height, and $\delta=(T^*-T_0)/T_0$, which measures the heat imparted, with proportionality constant  $\kappa_1 = 0.8$ for monatomic ideal gas ($c_v=3/2$, $\gamma=5/3$).     
One can similarly derive  small amplitude oscillations of the partition, at a characteristic frequency $\tau_s^{-1}$, or the sound speed divided by the characteristic size $L$.

\begin{figure}
	\centering
		\includegraphics[scale=0.27, angle=0]{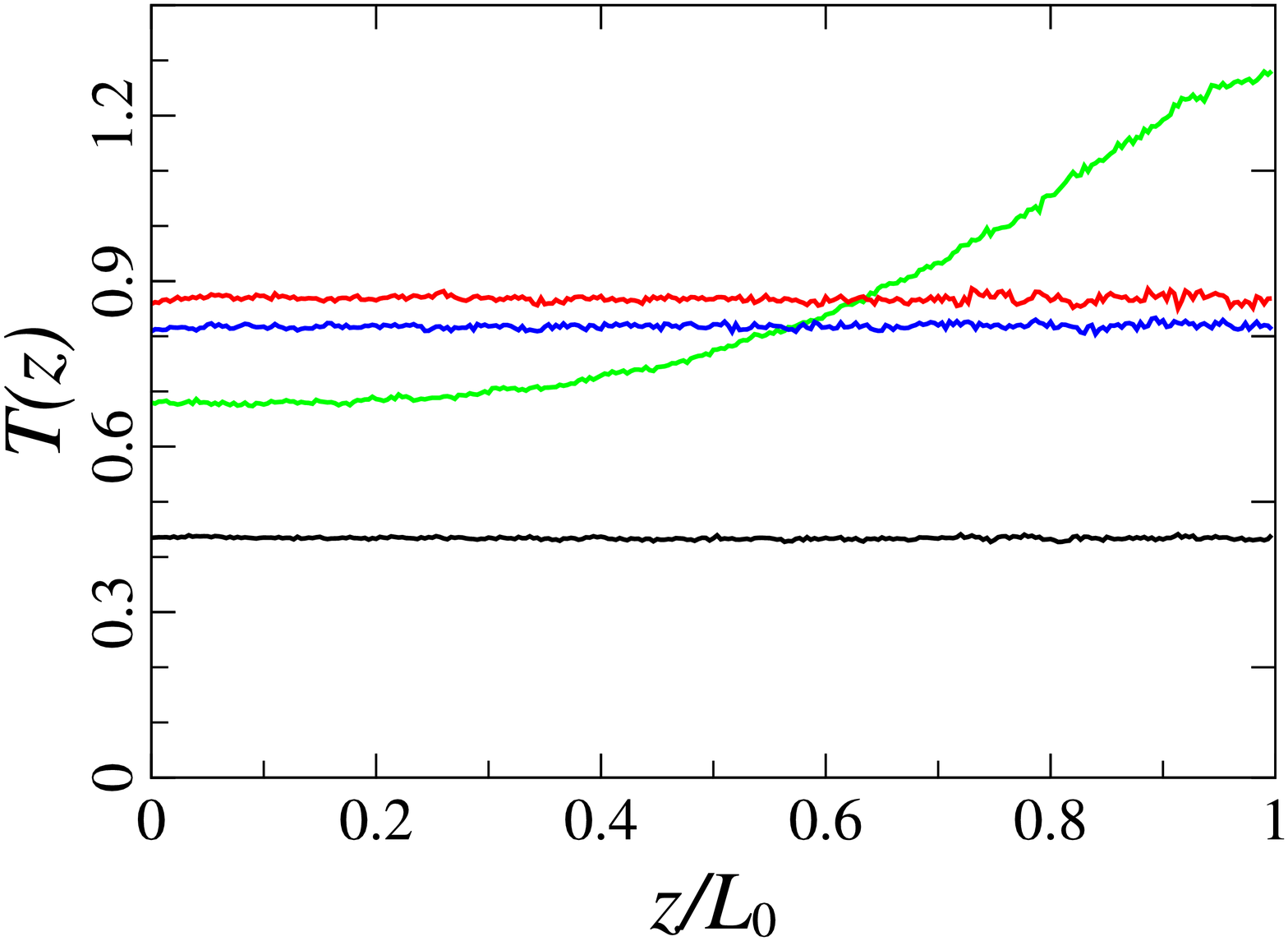}
	\caption{Temperature profile for fast heating. Black: initial profile; red: upon  heating; green: at maximum temperature gradient; blue: at new equilibrium.   ($\delta=G=1.0$.)}
	\label{fig:T_flat}
\end{figure}

\begin{figure}
	\centering
		\includegraphics[scale=0.27, angle=0]{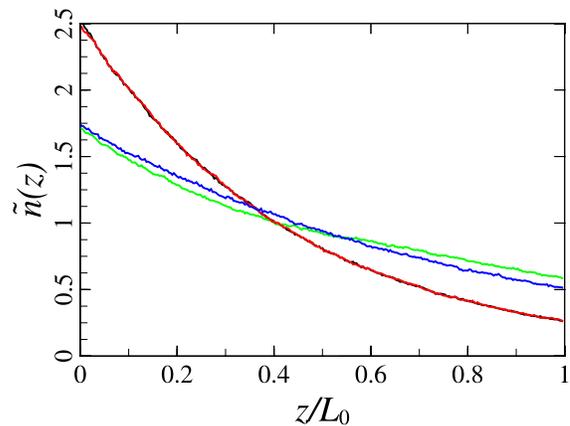}
	\caption{Density profile for fast heating.    Black: initial profile; red: upon  heating; green: at maximum temperature gradient; blue: at new equilibrium.   ($\delta=G=1.0$.)}   
	\label{fig:n_flat}
\end{figure}

\begin{figure}
	\centering
	\includegraphics[scale=0.26, angle=0]{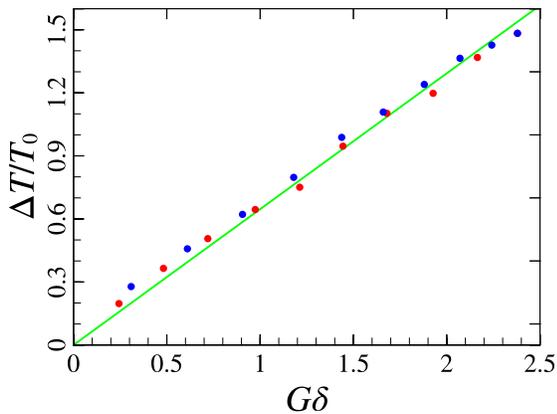}
	\caption{
	Bilinear dependence of normalized temperature differential of $\Delta T/T_0$ on  $G\delta$.
	Red dots:  $G =2.17$, $\delta$  varied from 0.11 to 1.1.
	Blue dots: $\delta =1.11$, $G$  varied from 0.28 to 2.16.
        Green line:  bilinear fit $y=0.64x$ to the numerical solutions.} 
	\label{fig:grad_amp}
\end{figure}

Numerical simulations confirm these effects also in the absence of a partition.        
In a vertical column, all particles receive a  lateral impulse proportional to their lateral velocity which, for hard-sphere collisions,   quickly  results in a vertically uniform temperature increase, whereupon a temperature differential develops, as depicted in Fig.~\ref{fig:T_flat}.  
The temperature differential is accompanied by a rise in the density scale height, as depicted in Fig.~\ref{fig:n_flat}.
The bilinear  dependence on  both $G$ and $\delta$ in Eq.~(\ref{eq:differential}) is confirmed, with  proportionality constant  $\kappa_1 \approx  0.64$ numerically determined, by fitting a large number of particle simulations  (see Fig.~\ref{fig:grad_amp}).  
The temperature gradient oscillates at the sound frequency.
As  the new equilibrium is reached, these oscillations are damped somewhat sooner than  the heat diffusion time, $\tau_H \approx  \nu_c L^2/v_t^2$,  for thermal speed $v_t$,  collision frequency $\nu_c = v_tnd^2$,  and  molecule radius  $d$.

Interestingly, in the limit  $L/ \lambda \rightarrow  \infty$, the temperature differential can no longer depend on $L$, since there are no particles at large $L$.  
Hence, the  bilinear dependence  derived in Eq.~(\ref{eq:differential}) must be replaced by a dependency  only on $\delta$. 
However, at the same time, the notion of temperature becomes ill-defined for too many scale heights.  
Focusing then on the temperature differential say between $z=0$ and $z=c \lambda$, where $c \equiv \ln 10$, one determines through simulations 
${\Delta T}/{T_0}=0.47 \delta$.
Note that the maximum temperature differential may occur at larger scale heights, but because of the increasing rare atmosphere, further contributions to the differential  become less meaningful with greater heights.

Note that in either limit the temperature differential can be of the order of the temperature increase itself.
To be concrete, consider for example a 100\,m high chamber,  filled with xenon gas at 273\,$^\circ$K.   
At standard  gravity,  a 10\,$^\circ$K  temperature differential results upon compression sufficient to heat  to 546\,$^\circ$K, 
assuming the compression time is less than the sound time of about 1\,s.   
The heat diffusion time is about 10$^9$\,s.
Greater differentials are possible with realizable centrifugal forces. 
Spinning xenon gas at 6000\,RPM in a cylinder with radius 10\,cm, 
and then heating it (like through axial compression) from T=273\,$^\circ$K to T=1355\,$^\circ$K, 
gives a temperature differential of 120\,$^\circ$K between the axis and the periphery. 
Note that this example is in the range of temperatures in combustion engines, where having the hot spot in the center may be advantageous in reducing wall losses.  
For the 10\,cm engine, the sound time is about 0.1\,s, making 6000\,RPM marginally in the range of validity of Eq.~(\ref{eq:differential}).  
The heat diffusion time (about 10\,s) is easily long enough.


{\it Slow Compression:} \ 
For compression times exceeding the sound time, 
namely  the case  $ \tau_c \ll \tau_s \ll  \tau_E  \ll \tau_H$,
one might speculate that, once a temperature differential develops upon compression, namely after some change in the scale height, 
further lateral compression accentuates that differential.
After all, compression heats at each height proportional to the local vertical temperature, so less energy will go to the cooler regions.  
On the other hand, slower heating means slower rise times, which would counteract this effect.
It turns out, however, that, except for lateral compression $x(t)$ specifically timed to exploit the oscillations in temperature, slower compression produces smaller temperature differentials.

To see this, note that compression  slow compared to sound time implies pressure balance as in \Eq{eq:F_bal};  for small partition displacement $z$, we have
\begin{gather}\label{eq:mean_z}
{z}=\frac{(\eta-1)Mg}{T_0\eta\left({N_1}/{L_1^2}+{N_2}/{L_2^2}\right)},
\end{gather}
where $\eta=(x(t)/x_0)^{1-\gamma}$.
From energy conservation, the energy required to compress may be put as 
\begin{equation}\label{eq:energy_c}
Q=Mg{z}+c_vN_1 (T_1-T_0)+c_v N_2 (T_2-T_0),
\end{equation}
whereupon solving for $z$, we have
\begin{equation}\label{eq:mean_z_m}
\begin{aligned}
{z}=\frac{MgQ}{T_0\left({N_1}/{L_1^2}+
{N_2}/{L_2^2}\right)\left(Q_0+Q\right)},
\end{aligned}
\end{equation}
where we defined $Q_0=c_v T_0(N_1+N_2)$.  Approximating $L_1=L_2=L/2$ and $N_1=N_2=N/2$, then gives  
\begin{equation}\label{eq:slow}
\begin{aligned}
\frac{\Delta T}{T_0}\equiv\frac{(T_2-T_1)}{T_0}=
(\gamma-1)G   \frac{Q}{Q_0} \equiv (\gamma-1)G \delta,
\end{aligned}
\end{equation}
which is now in the same bilinear form as Eq.~(\ref{eq:differential}),with approximately the same proportionality constant.  
However, by fitting numerical simulations of monatomic gas we find numerically a slightly smaller proportionality constant, namely $\Delta T/{T_0} =0.26\, G \delta$.
This indicates that unless a very specific compression time profile is taken,  compressing slower than the sound time gives a slightly smaller temperature differential compared to compressing faster.

\begin{figure}
	\centering
		\includegraphics[scale=0.28, angle=0]{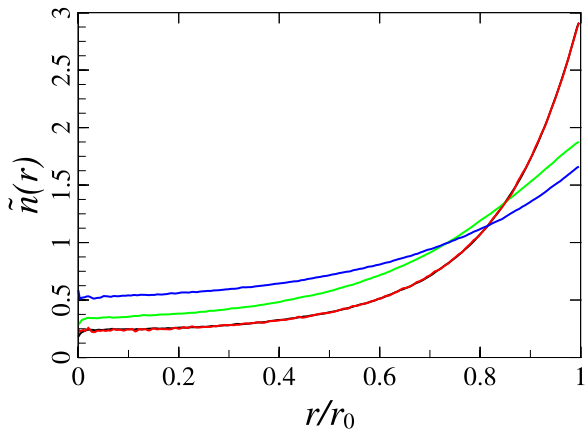}
\caption{Spinning gas density profile. Black:~initial profile; red:~upon heating; green: at maximum temperature gradient; blue: new equilibrium.  (Fast heating; $\delta=8/3$;  $\varphi_0=2.74$.)}
	\label{fig:n_circ}
\end{figure}
\begin{figure}
	\centering
		\includegraphics[scale=0.28, angle=0]{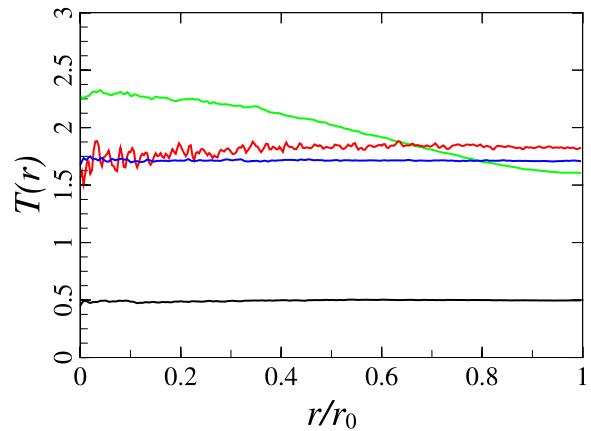}
	\caption{Spinning gas temperature profile.    Black: initial profile; red: upon heating; green:  at maximum temperature gradient; blue:  new equilibrium.   (Fast heating; $\delta=8/3$;  $\varphi_0=2.74$.)}
	\label{fig:T_circ}
\end{figure}

{\it Centrifugal Potential:} \
The piezo-thermal effect under a centrifugal potential is considerably more complicated than under a gravitational potential because the temperature is coupled to the centrifugal potential.  
However, by analogy to the gravitational potential case, it can be expected that the temperature differential will be bilinear in the heat imparted through compression and the initial centrifugal potential parameter $\varphi_0=m\omega^2r_0^2/2T$.
To see the analogous evolution effects, a spinning gas is simulated, with ignorable axial and azimuthal coordinates, and with specular reflection off the perfectly cylindrical boundary.  
The collision operator exploited the azimuthal symmetry to allow for hard-sphere collisions based on radial location only.  
The axial compression on a time scale short compared to a sound time is simulated by taking $v_z \to \alpha v_z$ for all particles, where $z$ is the axial direction. 
As in the gravitational case, the imparted energy  quickly becomes isotropic and thermally distributed, followed by radial density redistribution and a radial temperature differential.
The collisions insure that the system remains at all times very close to solid body rotation, even as the temperature and density evolve radially. 
Like in the 1D gravitational piezo-thermal effect, the temperature gradient in this limit oscillates several times before the new equilibrium is reached. 
In Fig.~\ref{fig:n_circ} the density evolution is shown.  
In Fig.~\ref{fig:T_circ}, the temperature differential forms very much like in the case depicted in Fig.~\ref{fig:T_flat}.
As anticipated,  the hotter region is near the axis.
Note, however, that the region close to $r=0$ is very noisy, since, for the centrifugal potential,  there are very few particles on axis.
Nonetheless, it can clearly be seen that a significant temperature differential is established. 

These simulations are also consistent with the \textsl{rotation-dependent heat capacity} effect \cite{ref:geyko_e}. 
Since neither temperature nor angular velocity are constant under compression, 
the spinning parameter $\varphi=m\omega^2r_0^2/2T$ varies in time. 
Hence, in transiting from one equilibrium to another, an average heat capacity  can be estimated. 
Here, the initial equilibrium has $\varphi=2.740$ at initial temperature $T_0=0.501$, while the final equilibrium (after compression) has $\varphi=1.470$ at temperature $T_f=1.233$.  
The energy per  particle in bringing the system from the initial state to the final state was $\Delta E=1.232$ (temperatures and energy here are dimensionless). 
The additional term for  heat capacity can be estimated using $(c_v+B)(T_f-T_0)=\Delta E$.
For monotomic ideal gas with $c_v=1.5$,  we find $B=0.168$, intermediate between the equilibrium values at the initial state, $B(\varphi_0)=0.305$, and at the  final state,  $B(\varphi_f)=0.119$ \cite{ref:geyko_e}, as expected since the spinning parameter changes under compression.

{\it Discussion:} \
Note that the piezo-thermal effect identified here produces an apparently opposite temperature gradient  from that found  in vortex tubes, which have the coldest flow on axis \cite{ref:liew,ref:gao,ref:ahlborn}. 
Although vortex tubes also exploit non-equilibrium effects,  the  temperature differentials occur through different mechanisms, such as by gas leaving the device through counterpropagating flows in the axial direction,  by turbulence, or by frictional forces.  

As a matter of fact, the piezo-thermal effect identified here can also produce cold flow on axis, since the effect changes sign with expansion rather than  compression.  
Thus, axial expansion of spinning gas, produces the coldest point on axis.
Similarly, lateral expansion of the gas in a gravitational field produces a temperature differential with the bottom hottest.

The piezo-thermal effect might also be run in reverse; a temperature gradient can produce contraction.  
Thus, imagine, as in Fig.~\ref{fig:gas_mass}, that the upper chamber is  heated in less than a sound time, but an equal amount of heat is taken from the bottom chamber.  
Then, initially, because the force is proportional to the total contained energy, there is no change in the force on the horizontal walls.   However,  to restore pressure balance,  the gas on bottom is compressed, so the partition is lowered, thereby releasing gravitational potential energy, and thus creating more net pressure on the horizontal walls.
The reverse effect, producing horizontal constriction, would occur by imposing the opposite temperature differential.

What is particularly fascinating about this piezo-thermal  effect is that it is generalizable to compression under any potential and to multiple gas constituents; we considered here both gravitational and centrifugal potential, but for a single gas constituent.  
To see the possibilities in more general cases, consider, for example,  that, under an electric potential, oppositely charged molecules experience opposite forces.
If the temperature equilibration between the species takes longer than the sound time, then this can lead to interesting effects. 
Consider then a column of plasma, in a vertical potential,
such as would occur in the region of a plasma sheath.
Upon lateral compression, a vertical temperature differential will develop, but, interestingly,  this differential will be opposite for electrons and ions.  
More generally yet, it is hoped that the identification and  elucidation here in special cases in a variety of limits  of the non-equilibrium piezo-thermal effect will lead to variations and elaborations on this effect in a variety of physical contexts.


{\it Acknowledgments} \
Work  supported by   DTRA,  DOE  Contract No.~DE-AC02-09CH11466, and by  NNSA SSAA  Grant No.~DE-FG52-08NA28553.

\end{document}